\documentclass[runinaddress,showpacs,twocolumn,prl]{revtex4}
\usepackage{amsmath}
\usepackage{amssymb}
\usepackage{graphicx}

\input{tcilatex}

\begin{document}

\title{Comment on "Nonexistence of "Spin Transverse Force" for a
Relativistic Electron" by Wlodek Zawadzki (cond-mat/0701387)}
\author{Shun-Qing Shen}
\affiliation{Department of Physics, and Center for Theoretical and Computational Physics,
The University of Hong Kong, Pokfulam Road, Hong Kong, China}
\date{January 19, 2007}

\begin{abstract}
This is a reply to W. Zawadzki's paper (arXiv: cond-mat/0701378) on
non-exietence of spin transverse force for a relativistic electron. The
force was first proposed by the present author that the spin current will
experience a transverse force in an electric field as a relativistic quantum
mechanical effect, and in semiconductor with Rahsba spin-orbit coupling.
Zawadzki's approach is based on an incorrect relation between the velocity
and canonical momentum, and his conclusion is not true.
\end{abstract}

\maketitle

Spin transverse force was proposed by the present author for electron spin
moving in an electric field in non-relativistic quantum mechanical limit or
in semiconductor with spin-orbit coupling.\cite{Shen05prl} The force
direction is perpendicular to the electric field and its amplitude is
proportional to spin current with polarization along the field. Some
physical effects such as the spin and anomalous Hall effect can be
attributed to this force.\cite{Li05prb} In his comment,\cite{Zawadzki07} W.
Zawadzki claimed that the force does not exist, and is an artefact resulting
from an approximate form of the employed Hamiltonian. He reproduced a
standard approach in the textbook on relativistic quantum mechanics, but
cited an incorrect relation between the canonical momentum operator and the
velocity. Thus his statement is not correct.

The complete Dirac Hamiltonian in the electric and magnetic field reads%
\begin{equation}
H=c\alpha \cdot (\mathbf{p}-\frac{e}{c}\mathbf{A})+\beta m_{0}c^{2}+V
\end{equation}%
where $\mathit{V}$ and $\mathbf{A}$ are the scalar and vector potential for
electromagnetic field, and $\alpha $ and $\beta $ are the Dirac matrices. $%
m_{0}$ is the rest mass of electron, and $c$ is the speed of light.
Zawadzki's starting point is based on the following equation,%
\begin{equation}
\mathbf{F}=\frac{d}{dt}(\mathbf{p}-\frac{e}{c}\mathbf{A})=e(\mathbf{E}+%
\mathbf{v}\times \mathbf{B}),
\end{equation}%
where $\mathbf{E}$ and $\mathbf{B}$ are the electric and magnetic field,
respectively, and $\mathbf{v}=c\alpha $ the velocity operators with
eigenvalues $\pm c.$ One can find a detailed derivation in Ref. \cite%
{Greiner2000} \ Though it has, at least, the form of the Lorentz force, its
expectation may not be useful because $\mathbf{v}$ contains the so-called 
\textit{Zitterbewegung }even for a free electron. It is clear that there is
no explicit equation of motion to set up any relation between the velocity
operator, $\mathbf{v}=c\alpha $ and the operator $\mathbf{p}-\frac{e}{c}%
\mathbf{A}$ in relativistic quantum mechanics. The most relevant operator to
the classical force is the expectation value of the acceleration $d\mathbf{v}%
/dt$ in classic limit. However, Zawadzki simply took $\mathbf{p}-\frac{e}{c}%
\mathbf{A}=m(v)\mathbf{v}$,\cite{Zawadzki07} where he claimed that $m(v)$ is
the Lorentz mass. Even in the absence of any electromagnetic potential, $%
\mathbf{p}=-i\hbar \nabla $ and is not proportional to the velocity $c\alpha
.$ For instance in a plane wave, $\mathbf{p}$ is a good quantum number, but
the plane wave may not be one of the four eigenstates for $\alpha $. Thus
even if the equality were valid $m(v)$ must be an operator, not a
coefficient. Unfortunately, the velocity operator has only two eigenvalues $%
\pm c,$ and the Lorentz mass $m(v)=m_{0}/\sqrt{1-v^{2}/c^{2}}$ becomes
divergent when the velocity reaches at the speed of light. Thus the relation
is not correct, and all his comments become groundless.

Indeed, the spin transverse force in Ref. [1] is derived in the
non-relativistic limit. In this approach, the four component Dirac equation
is reduced to the two component Pauli equation with several additional
terms. It is inevitable to introduce the spin-orbit coupling $\frac{e\hbar }{%
4m^{2}c^{2}}\sigma \cdot \lbrack E\times (\mathbf{p}-\frac{e}{c}\mathbf{A})]$
in the effective Hamiltonian, which term indeed does not appear in the Dirac
equation explicitly and can be regarded as quantum interference between the
electron and positron states, or majority- and minority component of Dirac
wave function. In the Pauli equation, all quantities are well-defined and
self-contained, and the relation between the velocity $d\mathbf{x}/dt$ and
the operator $\mathbf{p}-\frac{e}{c}\mathbf{A}$ is explicit by introducing
an additional spin gauge field. In this limit the electron mass is of course 
$m_{0}$, not the Lorentz mass as Zawadzki claimed. The relativistic
correction has been included in the effective Hamiltonian. Higher-order
terms missed in my derivation has slight revision on the final results. The
effective Hamiltonian with spin-orbit coupling for electron has been working
very well in condensed matter physics, and all relevant parameters can be
determined experimentally. We have no reason to drop this type of the
effective Hamiltonian in solid. The spin transverse force clearly originates
from the spin-orbit coupling, and is applicable to describe the
two-component wave function for electron. For a single spin the force may be
hard to be detected experimentally. However, the force is believed to be
measurable once the spin current circulates in semiconductors, not high
energy physics because the energy gap between the conduction band and
valence band in semiconductors is much smaller than the gap between the
electron and positron. Theoretically, the contribution from the minor
component of the Dirac wave functions has been already reflected in the term
of spin-orbit coupling. If insisting on starting from the so-called complete
Dirac Hamiltonian for semiconductors, we believe that it will have a quite
different description for electrons in semiconductors. No matter whether it
is applicable or not it is still practical to apply the theory of electronic
band structure to explore the physical properties in semiconductor.

Finally I would like to point pout that the concept of force is of classical
not quantum mechanics. It has physical meaning only when it was related to
physical observables to quantum system. For example, the spin dependent
force was first realized in the Stern-Gerlach experiment even before the
discovery of electron spin, which is proportional to $\sigma \cdot
\bigtriangledown \mathbf{B}$. To summarize, Zawadzki's statement is based on
an incorrect relation between the velocity and the operator $\mathbf{p}-%
\frac{e}{c}\mathbf{A}$. He mixed the physical quantities in four- and
two-component Hilbert space. Thus his statement is not true.

\end{document}